\documentstyle[prl,aps]{revtex}
\begin{document}

\title{Magnetic Field Effect in Josephson Tunneling between $d$-Wave Superconductors}

\author{Xin-Zhong Yan$^{1,2}$ and Chia-Ren Hu$^1$}

\address{$^1$Department of Physics, Texas A\&M University, College Station, TX 77843-4242 \\
$^2$Institute of Physics, Chinese Academy of Sciences, P. O. Box 603, Beijing 100080, China} 

\date{\today}

\maketitle

\begin{abstract}
The magnetic field effect in the Josephson tunneling between two d-wave superconductors are investigated. When the crystal orientation of one (or each) superconductor relative to the interface normal is such that midgap states exist 
at the interface, there is a component of the tunneling current due to the midgap states.  For a junction with a flat \{001\}$|$\{110\} or \{100\}$|$\{110\} interface, this component is the predominant contribution to the current. The predicted current-field dependence differs entirely from the conventional 
Fraunhofer pattern, in agreement with a published measurement.  This is because, apart from the Fraunhofer factor, the critical current depends on the magnetic field $B$ through the current density also which is a linear function of $B$ for weak $B$.
\end{abstract}

\twocolumn
Whether the magnetic-field dependence of the Josephson tunneling current between two high-$T_c$ superconductors (HTSC's) should show the well-known Fraunhofer pattern seen in junctions made of conventional superconductors is still an unsettled question. There seem to be no theoretical discussions sheding light on this subject so far.  On the other hand, many measurements have been devoted to the investigation of this problem [1].  However, it is not easy to fabricate high-quality junctions of HTSC's with good flat interfaces.  Since the pairing order parameter in HTSC's very likely has predominately $d$-wave symmetry, the behavior of the Josephson tunneling current is expected to depend strongly on the interface orientations.  Most of the Josephson junctions exhibit random patterns in their current-field relations most-likely because their interfaces are made of many facets with a variety of orientations [1,2]. Recently, Ishimaru {\it et al.} have successfully fabricated junctions with atomically flat \{001\}$|$\{110\} interfaces [3]. Their measurements showed a regular dependence of the Josephson critical current on the applied magnetic field, which is fundamentally different from the Fraunhofer pattern.  Such a seemingly anomalous behavior of the Josephson current has not yet been explained in an acceptable theory. 

In this letter, we present a calculation of the Josephson tunneling current between two $d$-wave superconductors (DWSC's) with a \{001\}$|$\{110\} interface, as is investigated in that experiment. Our calculation shows that the local 
Josephson-current density depends linearly on the magnetic field when the field is weak.  Therefore, the current-field relation in such a Josephson tunneling junction is predicted to be different from the usual Fraunhofer pattern. This behavior stems from the existence of the midgap states (MS's) at the junction interface [4-8].

It has been noted that the MS's play a very important role in Josephson tunneling processes between two DWSC's [9] (and also in quasi-particle tunneling in junctions involving DWSC's which does not concern us here).  Consider a SC with a \{110\} surface at $x=0$, with the $x$ axis being defined perpendicular to this surface.  Let a magnetic field $\vec B$ be applied along the $c$-axis, so that there is a vector potential $\vec A$ along [1$\bar 1$0], which is designated as the $y$ axis. In the gauge in which the order parameter is real, the vector potential vanishes in the interior of the SC.  We have $A(\vec r) \simeq -B\lambda \exp(-x/\lambda)$, where $\lambda \simeq 1500\AA$ is the magnetic-field penetration depth in the $ab$ plane. The perturbation to the motions of particles is described by the Hamiltonian,
$$H' = -{1\over c}\int d^3r\,{\vec j}(r)\cdot{\vec A}(r).  \eqno(1)$$
where $c$ is the speed of light, and ${\vec j}(r)$ the current density. The particles in the continuum states should only be perturbed very weakly by this magnetic field since their wave functions extend far away from the surface where $B\simeq 0$. In contrast, the perturbation to the MS's is significant because the MS's are localized within roughly a coherence length ($\xi\sim 15\AA$) from the surface, which is much smaller than $\lambda$. From the Bogliubov-de Gennes equations for quasi-particles which are coherent superpositions of electrons and holes, the eigen-energies of the perturbed MS's can be obtained as
$$\varepsilon_k \simeq {\hbar e\over mc}A(0)k = -{\hbar e\lambda\over mc}Bk, \eqno(2)$$
where $k$ is the wave number along $y$, and we have neglected the Zeeman energy. [10]  This energy expression is an odd function of $k$. 
Thus at sufficiently low temperatures, an asymmetric distribution of quasi-particles in the MS's can be realized. 

In the Josephson junction under consideration, the wave functions of the MS's are distributed predominately in the side with a \{110\} surface, with only a small amplitude tunnelling into the other side (assuming high tunneling barrier at the interface).  These states are Andreev bound states localized within roughly a coherence length from either side of the interface. It is well known that 
quasi-particles in such bound states can transport electric current between two superconductors. In the elementary charge-transport process in this problem, the direction of the tunneling current mediated by such a state will depend on the sign of its momentum $k$ along $y$, since this tunneling current is proportional to the wave-function product $u_k^*v_k$ which is proportional to the order parameter of this SC at its surface, the sign of which depends on the sign of $k$.  At zero magnetic field two states of momentum $k$ and $-k$, respectively, have the same energy $E(k) = E(-k)$.  They are thus equally occupied, so that their contributions to the Josephson current cancel each other.  Under a small non-vanishing magnetic field, the energies of two MS's of such a pair of $k$ values as given by Eq. (2) are opposite in sign.  Their occupations at low temperatures will no longer be equal.  With cancellation no longer complete, a net current can now be transferred across the Junction via this pair of MS's which is proportional to the difference of their occupation $f(\varepsilon_{k}) - f(\varepsilon_{-k}) = \tanh(\varepsilon_{-k}/2k_BT)$, where $f(\varepsilon_k)$ is the Fermi distribution function at temperature $T$ and $k_B$ is the Boltzmann constant. If the magnetic field applied in the tunneling experiment is so weak that $|\varepsilon_k| \ll 2k_BT$ --- roughly, the ratio $|\varepsilon_{k}|/2k_BT$ is less than $0.05 B/T$ where $B$ and $T$ are in units of Gauss and Kelvin, respectively, --- the current density $J(B)$ is then a linear function of $B$. The total critical current $I_c(B)$ of a Josephson junction is determined by this current density times a factor describing a Fraunhofer-like pattern.  As the final result, at weak magnetic field, $I_c(B)$ is totally different from the usual Fraunhofer pattern.

To quantitatively study this problem, we use the tunneling-Hamiltonian approach in a tight-binding model. For the sake of description, the SC side with a \{110\} (\{001\}) surface is referred to as the ridgt (left) side of the junction.  The bridge width of the junction is denoted as $L$.  As is the situaton in the experiment, we assume that the barrier thickness and the length of the left SC in the direction perpendicular to the interface are both much less than $\lambda$. Then the effects of the magnetic field in this side of the junction and in the barrier region can be neglected.  Note that the external magnetic field $B_{ext}$ applied to the junction can induce a paramagnetic field $B_{para}$ to the right of the surface region occupied by the MS's in the right SC, so that the total magnetic field applied to the right SC is practically $B = B_{ext}+B_{para}$, which is then screened in a distance $\sim\lambda$ in the right SC.  According to the tight-binding model adopted here, the electrons are supposed to tunnel from the surface atoms of one SC to the neighboring atoms on the surface of the other, through the interface.  The Hamiltonian describing this tunneling process is given by
$$H_T = \sum_{ij\alpha}(T_{ij}d^{\dagger}_{i\alpha}c_{j\alpha} + H.c.) \eqno(3)$$
where $d^{\dagger}_{i\alpha}$ ($c_{j\alpha}$) is the creation (annihilation) operator for electrons with spin-$\alpha$ in the right (left) SC, the $i$ and $j$ summations run over the surface sites on the two sides of the junction, respectively.  (In the actual calculation, each sum is over one infinite chain of atomic sites only, since for the right side we need only consider one CuO$_2$ plane with one linear \{11\} edge, which practically only interacts with one linear \{10\} chain of atoms on the left side.) The tunneling matrix element is then $T_{ij} = |T_{ij}|\exp[i(\phi_0/2+k_0y_i)]$, with $k_0=eB\lambda/\hbar c$ and $\phi_0$ is the initial phase difference between the two SC's.  The magnitude of the tunneling matrix element $|T_{ij}|$ is considered as a function of the distance $|y_i-y_j|$ (where $y_i$ and $y_j$ are respectively the coordinates of the sites $i$ and $j$ along the interface). Here, we suppose a simple functional form $|T_{ij}| \equiv T(y_i-y_j) = T_0\exp(-(y_i-y_j)^2/r^2_0)$ with the parameter $r_0$ describing the effective tunneling range. (The precise functional form is not important as long as it is short-ranged. The chosen form is for convenience only.) By the standard perturbation treatment, the Josephson tunneling current can be expressed as 
$$I = 2e{\rm Im}\Phi_r(\omega)|_{\omega=0}  \eqno(4)$$
with $\Phi_r(\omega)$ the analytical continuation $i\omega_n$ (= $i2n\pi k_BT$) $\to \omega+i 0^{+}$ of the correlation function $\Phi_r(i\omega_n)$. In terms of the one-particle Green functions, 
$$\Phi_r(i\omega_n) = -2k_BT\sum_{ii'jj'}\sum_{\ell}T_{ij}T_{i'j'}
                F^r_{i'i}(\Omega_{\ell})F^l_{jj'}(\Omega_{\ell}+\omega_n) \eqno(5)$$
where $\Omega_{\ell}=(2\ell+1)\pi k_BT$ is the Matsubara frequency, and 
$F^r_{i'i}(\Omega_{\ell})$ and $F^l_{jj'}(\Omega_{\ell})$ are the Fourier transforms of the imaginary-time ($\tau$) anomalous Green functions $F^r_{i'i}(\tau) = -<T_{\tau}d^{\dagger}_{i'\downarrow}(\tau)d^{\dagger}_{i\uparrow}(0)>$ and $F^l_{jj'}(\tau) = -<T_{\tau}c_{j\uparrow}(\tau)c_{j'\downarrow}(0)>$, respectively. Along the interface of the junction, the anomalous Green functions depend on the relative distance of two $y$-coordinates only.  Accordingly, we factorize $T_{ij}T_{i'j'}$ to $|T_{ij}T_{i'j'}|\exp[ik_0(y_{i'}-y_i)]$$\exp[i(\phi_0+2k_0y_i)]$.  The factor $\exp[ik_0(y_{i'}-y_i)]$ may be approximated by unity, since it leads to a negligible correction to a momentum conservation relation obtained later when Fourier transform is employed. ($k_0$ is estimated to be $\simeq$228 cm$^{-1}$ at 
$B = 1G$, whereas the important momenta are all of order $k_F$.)  The summation over $i$ which can be performed immediately then produces the factor describing a Fraunhofer-like pattern: $j_0(k_0L) = \sin(k_0L)/k_0L$. 

To illustrate the physics, we neglect the space variation of the order parameter near the interface since the existence of MS's is invariant with respect to such variations. With this simplification, the wave functions of the MS's and also the Green's functions can be obtained analytically within the tight-binding model defined in a square lattice [8,11]. In momentum space, the anomalous Green's function $F^l$ is given by
$$F^l(k,\Omega_{\ell}) = 
 -a\int^{\pi/a}_{-\pi/a}{dq_x\over 2\pi}{\Delta_q\over \Omega^2_{\ell}+E^2_q}, \eqno(6)$$
where $a$ is the lattice constant, $\Delta_q = 2\Delta[\cos(q_xa)-\cos(ka)]$ and $E_q$ [with $q \equiv (q_x, k)$] are the order parameter and the quasi-particle energy, respectively [8,10]. For the problem under consideration, we need only consider the current transferred via the MS's. Therefore, the anomalous Green function $F^r$ is simply:
$$F^r(k,\Omega_{\ell}) = {u^{*}_kv_k\over i\Omega_{\ell}-\varepsilon_k}, \eqno(7)$$
where $u^{*}_k$ and $v_k$ are the surface values of the wave functions of the MS with momentum $k$. In the lattice model, the eigen-energy of a MS is given by $\varepsilon_k = -(2{\sqrt 2}ateB\lambda/\hbar c)\sin (ka/\sqrt{2})$, and $t$ is the hopping energy of electrons between nearest-neighbor sites. It is worth noting that $v_k = iu_k{\rm sgn}k$ for these MS's, which means that the tunneling-current direction depends on the sign of $k$ as mentioned above. Also, due to the phase factor $i$, the phase dependence of the Josephson current is $\cos \phi_0$ rather than the usual $\sin \phi_0$ . Substituting these results into Eq. (4), we obtain 
$$I = J(B)j_0(k_0L)\cos\phi_0, \eqno(8)$$
$$J(B)= eN_s\sum_{n=-\infty}^{\infty}\int^{\pi/{\sqrt 2}a}_0\,
                        {dk\over\pi}\,T^2(k+Q_n)\chi(k,Q_n), \eqno(9)$$
$$\chi(k,Q_n)= |u_k|^2\int^{\pi/a}_{-\pi/a}\,{dq_x\over 2\pi}\,
{\Delta_q\over E_q}[g(\varepsilon_k,E_q)-g(\varepsilon_k,-E_q)], \eqno(10)$$
where $N_s$ is the number of total lattice sites on the \{110\} surface of the right SC, $T(k)$ is the Fourier transformation of $T(y)$, $q = (q_x, k+Q_n)$ with $Q_n = 2n\pi/{\sqrt 2}a$, and $g(\varepsilon_k,E_q) = 
[\tanh(\varepsilon_k/2k_BT)-\tanh(E_q/2k_BT)]/(\varepsilon_k -E_q)$.  Apart from the factor $j_0(k_0L)$, the Josephson current depends on $B$ through $J(B)$ also. At weak $B$, $J(B) \propto B$. This is an unusual feature of the Josephson current between two DWSC's when MS's are present at the interface and give the dominant contribution to the current.  

In Fig. 1, we show the numerical results for $J(B)$ normalized by 
$J_0 = eN_sT^2_0\Delta^2/t^3$ at $T = 4.2 K$ for a junction with a \{001\}$|$\{110\} interface.  The long-dashed, solid, and short-dashed curves correspond to the parameters $q_0 \equiv r^{-1}_0 = 1.0a^{-1}$, $0.94a^{-1}$, and $0.9a^{-1}$, respectively. Obviously, $J(B) \propto B$ at weak $B$.  The slope of $J(B)$ at $B = 0$ depends sensitively on the parameter $r_0$. For a narrow region in the vicinity of $q_0 = 0.94a^{-1}$, $J(B)$ changes sign at a certain value of $B$ of the order of $B_0$.  This is because right at the interface, the pairing amplitude in the left SC is proportional to $\Delta\cos(ka)$.  Its sign and therefore the sign of the corresponding contribution to the tunneling current density depend on the magnitude of the momentum $k$ of the quasi-particle: $\cos(ka) \buildrel{>}\over{<} 0$ for $ka \buildrel{<}\over{>} \pi/2$. Owing to the $B$-dependence of the energy of a MS, $|\varepsilon_k| \propto B\sin(ka/\sqrt 2)$, changing magnetic field is equivalent to modulating the distribution of the quasi-particles in the $k$ space.  Similarly, the relative weights of positive and negative currents are also controled by $r_0$.  These effects result in the delicate dependence of $J(B)$ on $B$ and $r_0$.   

There is some subtlety in the factor $j_0(k_0L)$ as well:  In an ideal system, there exists a spontaneous current near the \{110\} surface of the right SC [10,12,13], so a $B_{ext}$-independent spontaneous paramagnetic field $B_{para} \equiv B_0 = \hbar c /2e\lambda^2$ (which is $\sim$146 Gauss as estimated by using $\lambda \simeq 1500$ $\AA$) is produced. The total magnetic field $B = B_{ext} + B_0$ in the right SC is nonzero even at $B_{ext} = 0$. Therefore, the factor $j_0(k_0L)$ alone can already cause the $I_c$-$B_{ext}$ curve to not have the usual Fraunhofer pattern.  In particular, there may be no main peak at $B_{ext}$ = 0 as in the usual Fraunhofer pattern. Generally, there is no unique pattern if this spontaneous field exists: At $B_{ext} = 0$, there may be a dip or a cusp in $I_c(B_{ext})$ depending on the geometric ratio $L/\lambda$ of the junction.  However, the existence of this spontaneous magnetic field depends sensitively on the quality of the juction including the flatness of the interface and the impurity content of the SC which has the MS's.  We do not believe that most of the HTSC's can have this spontaneous current.  Even for a perfectly flat \{110\} interface, the existence of impurity scatterings is unavoidable. In such a non-ideal system, the particles in the MS's have finite lifetimes due to their scatterings off the impurities.  Therefore, the spontaneous surface current and thereby the spontaneous magnetic field cannot exist. The induced paramagnetic field is no longer a constant but depends linearly on the external field, $B_{para} \propto B_{ext}$, for small field, when the magnitude of 
the maximum shift of the MS's is still much smaller than the width of the midgap-state peak in the density of states.  Consequently, the dependence of $j_0(k_0L)$ on the external field in the weak field regime is essentially the conventional Fraunhofer pattern except for a scale factor ($<1$) on $B_{ext}$. 

Shown in Fig. 2 is the result for $I_c(B)$ normalized by $J_0$ for the junction with $L/\lambda = 35$ and $q_0 = 0.94a^{-1}$. To compare with the experimental measurements, we need to take into account a larger enhancement factor in the effective magnetic field, which is mainly due to the flux focusing effect in thin-film grain boundary junctions [14].  Due to this effect, the effective field $B$ can be an order of magnitude larger than the external field $B_{ext}$. Under this consideration, the experimental result of Ishimaru {\it et al.} [3] can be explained by the present theory, as is shown in Fig. 2.  According to the present theory, the deep dip in the measured $I(B)$ curve at zero magnetic field is a strong indication that the pairing has predominately $d$-wave symmetry in the CuO$_2$ planes of YBCO. 

If the SC's have the $d+s$ symmetry, then the tunneling current can also be transferred through the continuous states.  In this case, the Josephson critical current $I_c(B)$ does not vanish at $B = 0$. There may appear a dip or a peak in the $I - B$ curve at $B=0$ depending on the magnitude of the $s$-wave component. The fact that the observed $I_c(0)$ is small with no peak there should imply that the $s$-wave component is small, though possibly not zero in YBCO.  The predicted features of the $I_c-B$ curve can also be observed in junctions with \{100\}$|$\{110\} interfaces if only they can be fabricated with flat interfaces. The physics is essentially the same as for the case of a \{001\}$|$\{110\} interface. 

In summary, we have investigated the magnetic-field dependence of the Josephson current in a junction made of two DWSC's with a \{001\}$|$\{110\} interface.  In such a junction the Josephson current is mainly mediated by the MS's on the side with a \{110\} surface. Under a magnetic field, the eigen-energies of these states shift in direct relation to their momenta along the interface, leading to a skewed occupation of these states, which implies a $B$-dependent current-density factor $J(B)$ which is linear in $B$ for small $B$. When the particles in the MS's have a finite life-time, the effective field $B$ acting on the side with a \{110\} surface should be a linear function of the external field when the field is weak. Thus the usual Fraunhofer factor is not modified qualitatively, but the critical current $I_c$ depends on $B$ through a current-density factor $J(B)$ also which varies linearly with $B$ at weak $B$. Thus the $I_c-B_{ext}$ curve differs entirely from the conventional Fraunhofer pattern, especially near zero field. 

This work is supported by the Texas Higher Education Coordinating Board under the grant No. 1997-010366-029, and by the Texas Center for Superconductivity at the University of Houston.


Figure Captions
\begin{itemize}

\item[Fig. 1]
The calculated current density $J(B)$ of Josephson tunneling between two $d$-wave superconductors with a \{001\}$|$\{110\} interface as a function of the magnetic field $B$ at $T = 4.2 K$.  The long-dashed, solid, and short-dashed lines correspond to the parameter $q_0 \equiv r^{-1}_0 = 1.0a^{-1}$, $0.94a^{-1}$, and $0.9a^{-1}$, respectively. The magnetic field is normalized by $B_0 = \hbar c /2e\lambda^2 $. 

\item[Fig. 2]
The calculated critical Josephson current $I_c(B)$ as a function of the magnetic field $B$ for the junction of two $d$-wave superconductors with a \{001\}$|$\{110\} interface at $T = 4.2 K$ and $q_0 = 0.94a^{-1}$. The ratio between the bridge width and magnetic field penetration depth (in the $ab$ plane) is $L/\lambda = 35$. The black dots are data from Ref. 3.

\end{itemize}

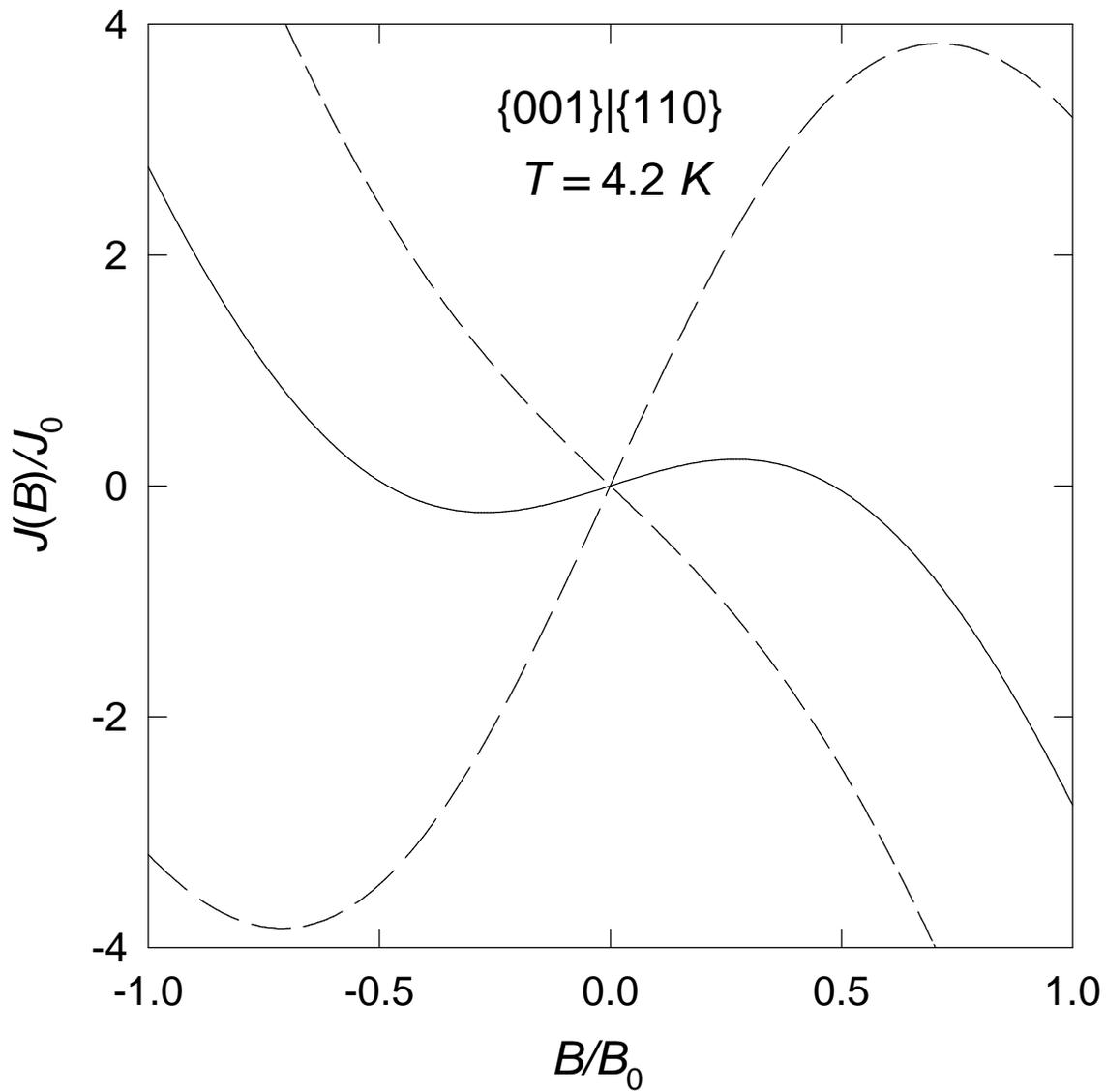
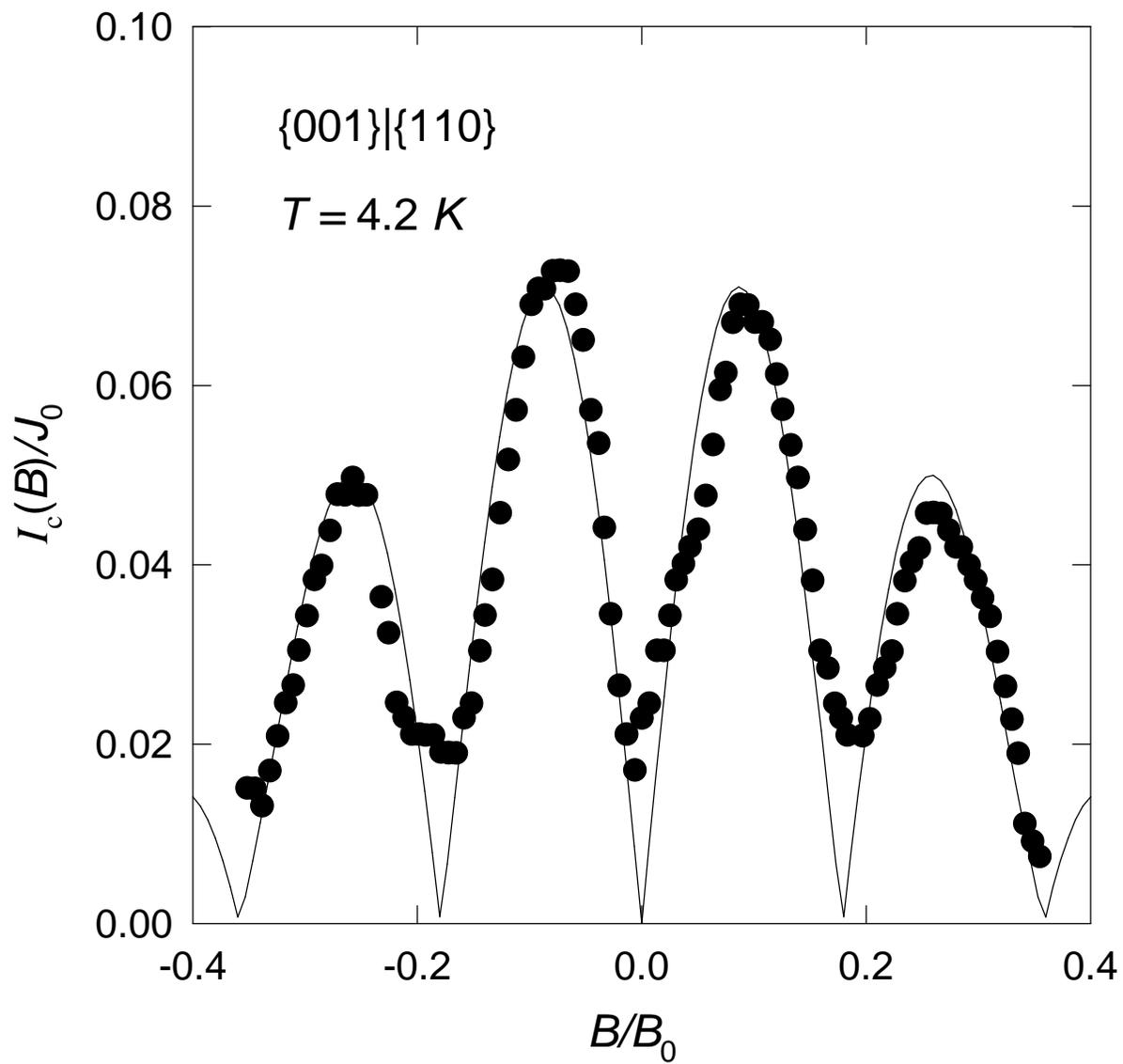
\end{document}